\documentclass[prl,superscriptaddress,twocolumn,showpacs,amsmath,amssymb,floatfix,preprintnumbers,nofootinbib]{revtex4}
\usepackage{epsfig}

\newcommand{\beqn}{\begin{eqnarray}}
\newcommand{\eeqn}{\end{eqnarray}}
\newcommand{\eq}[1]{(\ref{#1})}
\newcommand{\avr}[1]{\langle #1 \rangle}
\newcommand{\bp}{{\mathbf p}}
\newcommand{\dD}{{\mathrm D}}
\newcommand{\cD}{{\mathcal D}}
\newcommand{\cZ}{{\mathcal Z}}
\newcommand{\SB}{{\mathrm {SB}}}

\newcommand{\dd}{{\mathrm d}}
\newcommand{\Tr}{{\mathrm{Tr}}\,}
\newcommand{\Z}{{\mathbb Z}}
\newcommand{\intp}[1]{\int \hskip -1mm \frac{\dd^3 #1}{(2 \pi)^3}}

\newcommand{\logo}{\\ \vskip -18mm \leftline{\includegraphics[scale=0.3,clip=false]{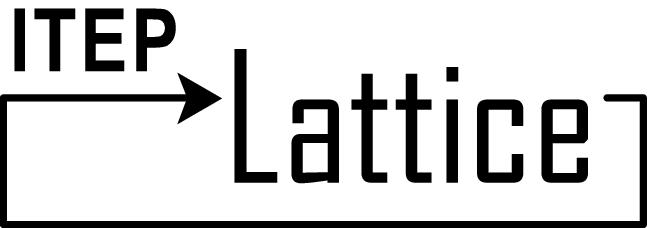}} \vskip 12mm}
\newcommand{\Hiroshima}{\affiliation{
Research Institute for Information Science and Education,
Hiroshima University, Higashi-Hiroshima, 739-8527, Japan}}
\newcommand{\Moscow}{\affiliation{Institute for Theoretical and
Experimental Physics, B.Cheremushkinskaya 25, Moscow, 117218, Russia}}
\newcommand{\Pisa}{\affiliation{
INFN -- Sezione di Pisa, Dipartimento di Fisica Universita di Pisa, Largo Pontecorvo 3, 56127 Pisa, Italy}}
\newcommand{\Munich}{\affiliation{Max-Planck Institut f\" ur Physik,
F\" ohringer Ring 6, 80805 M\" unchen, Germany} }

\begin{document}

\title{Combining infrared and low-temperature asymptotes in Yang-Mills theories\logo}

\author{M.N.~Chernodub}\Moscow\Hiroshima
\author{V.I.~Zakharov}\Moscow\Pisa\Munich

\begin{abstract}
We demonstrate that power-like non-perturbative behavior of gluon and ghost propagators in
infrared  limit of Yang-Mills theories can provide at finite temperatures $T$ a negative $T^4$--contribution to
the pressure and energy density. Existence of mass gap implies then
new relations between infrared critical exponents of
gluon and ghost propagators.
\end{abstract}

\preprint{ITEP-LAT/2007-04}
\date{December 14, 2007}

\maketitle

Zeros of Green's functions play important role in various condensed matter systems ranging from
superfluid ${}^3$He to unconventional superconductors~\cite{ref:Volovik}.
Under certain circumstances poles of Green's functions in momentum space -- which
determine the excitation spectrum of the system -- can undergo transition to their opposites,
that is, zeros. A known example of such a system is  the Mott insulator. In this case, a pole of
a one--particle Green's function becomes a zero as the system passes through the Mott
transition~\cite{ref:zeroes}. In field theory, a similar example is provided by
 strong interactions. As was first argued by Gribov \cite{ref:Gribov},
in the Landau gauge a perturbative pole of the gluon propagator is converted into a zero
at the vanishing momentum \cite{ref:FMR,ref:Zwanziger}.
This phenomenon is confirmed by both analytical calculations and numerical
simulations~\cite{ref:infrared:review}. The effect is related to the violation
of reflection positivity and is a signature of the color confinement.
The aim of this Letter is to demonstrate that anomalous behavior
of the gluon propagator in the infrared region is directly related to the
anomalous contribution of massless degrees of freedom at vanishing temperatures.
The very existence of a mass gap in confining theories constitutes then a new kind of
analyticity which constrains infrared asymptotes of the propagators.

We consider pressure $P$ and energy density $\epsilon\equiv E/V$
of strongly interacting Yang-Mills fields at temperature $T$ and in volume $V$.
In the limit of vanishing coupling, or in the tree approximation:
\beqn
\varepsilon_{\mathrm{free}}(T)  = 3 P_{\mathrm{free}}(T) = N_{d.f.} \, C_{\SB} \, T^4\,,
\label{eq:SB}
\eeqn
where $C_{\SB} {=} \pi^2/30$ is the Stefan-Boltzmann factor.
There are $N_{d.f.} = 2 (N_c^2 - 1) $ degrees of freedom corresponding to
$N_c^2 - 1$ gluons with two transverse polarizations.

Non-perturbative strong interactions are manifested in suppression of the energy
density and pressure at $T \ll T_c$, where $T_{c}$ is the temperature of the
confinement-deconfinement phase transition:
\beqn
\frac{\varepsilon(T)}{T^4} \ll 1\,,
\qquad
\frac{P(T)}{T^4} \ll 1
\quad \mbox{for} \quad
T \ll T_c\,.
\label{eq:suppression}
\eeqn
In particular, such a suppression is established via lattice simulations both
in the pure gluonic $SU(3)$ case~\cite{ref:Karsch:SU3} and in a more realistic
case which incorporates light quarks as well~\cite{ref:Karsch:quarks}.
Hereafter we will concentrate on the  pure gluonic $SU(3)$ gauge system.

The suppression~\eq{eq:suppression} is a consequence of two intimately related
properties of QCD, namely, confinement of color and mass-gap generation.
Massless gluons are confined into glueballs which are massive.
As a result, all the thermodynamic quantities are to be suppressed at low temperatures as $O(\exp\{- M/T\})$,
where $M$ is the mass of the lightest glueball, for a related discussion see~\cite{ref:Appelquist}.

The simplest quantity for a homogeneous system in the thermodynamic equilibrium is pressure,
\beqn
P = \frac{T}{V} \log \cZ = P_{\mathrm{free}} + P_{\mathrm{int}}\,
\label{eq:pressure:1}
\label{eq:pressure:sum}
\eeqn
where $Z$ is the partition function and we explicitly separate
the tree level contribution $P_{\mathrm{free}}$
given by Eq.~\eq{eq:SB}.
The second term $P_{\mathrm{int}}$ is a correction due to the interactions.
The interaction part of the thermodynamic quantities can be written as a loop expansion,
in terms of the full propagators and vertices, see, e.g., Ref.~\cite{ref:McLerran}.

In order to make explicit calculations one generically
needs a gauge fixing, and below we usually refer to the Landau gauge which -- being both
Lorentz--  and color--symmetric -- is one of the best studied gauges nowadays.
However, our  approach is not limited to the Landau gauge, and, as we will see below,
is in fact very general.
As a consequence, the final result -- a relation between the infrared exponents of
the dressed gluon propagator and the dressed ghost propagator should be gauge invariant,
provided that the infrared exponents (which may vary from gauge to gauge) are defined
as we discuss below.

Before going into details we would like to make a general comment about
the gauge invariance. Despite
of the fact that Green's functions in non-Abelian theories are essentially gauge-variant,
their properties may
be directly linked to the confinement. The best known
example of this kind is provided by Ref.~\cite{ref:Gribov} which relates the linear
confining potential increase of the
color charge interaction  at large distances to cancelation of the infrared pole of the gluon Green's function.

At finite temperatures, the gluon propagator $D^{ab}_{\mu\nu}$ and the ghost
propagator $G^{ab}$ in the Landau gauge are parameterized by
three functions $D_{{T}}, D_{L}$, and $D_G$:
\beqn
D^{ab}_{\mu\nu}(\bp,p_4) & = & \delta^{ab} \Bigl[P^T_{\mu\nu}\, D_{T}(\bp,p_4)
+ P^L_{\mu\nu}\, D_{L}(\bp,p_4) \Bigr]\,, \ \
\label{eq:propagator:Tneq0}\\
G^{ab}(\bp,p_4) & = & - \delta^{ab} D_G(\bp,p_4)\,,
\eeqn
where $P^T_{\mu\nu}$ and $P^L_{\mu\nu}$ are projectors onto spatially transverse and
spatially longitudinal parts~\cite{ref:Kapusta}.

A conspicuous feature of the propagators is the anomalous dressing of
the gluon and ghost propagators in the infrared region~\cite{ref:infrared:review}.
In the low temperature limit:
\beqn
D_i(p^2) & = & {(p^2)}^{\gamma_i - 1} \, H_i(p^2)\ \ \mbox{as} \ \ p^2 \to 0 \ \ ,
\label{eq:propagator:critical}
\eeqn
where $\gamma_i$ are infrared exponents and the functions $H_i$ are finite
at $ p^2 \to 0 $ in all the cases, $i=T,L,G$.

The infrared behavior of the propagators is best studied at zero temperature.
First of all, the gluon infrared exponents are degenerate, $\gamma_D = \gamma_T \equiv \gamma_L$.
We have already mentioned the Gribov's scenario for $\gamma_{D}>0$
\cite{ref:Gribov,ref:FMR,ref:Zwanziger}. Moreover, according to the Kugo-Ojima criterion
of confinement \cite{ref:Kugo:Ojima} the ghost propagator is enhanced in the infrared,
compared to its perturbative value:
\beqn
\gamma_D + 2 \gamma_G = 0\,.
\label{eq:KugoOjima}
\eeqn
There is some support for the validity of this relation, both from the analytical
studies of the Dyson-Schwinger equations
\cite{ref:Dyson:Schwinger,ref:infrared:review}
and from numerical, lattice data~\cite{ref:numerical,ref:large:volume2}.
However,  final conclusion on the validity of Eq. (\ref{eq:KugoOjima})
seems to be not yet reached,
see in particular Refs.~\cite{ref:analytical:against,ref:Dyson:Schwinger,ref:Michael,ref:CM}.
A cautionary remark comes also from a confining Abelian
gauge model  where one can prove analytically~\cite{ref:Suzuki}
that Eq.~\eq{eq:KugoOjima} does not hold.

Turning to non-zero temperatures, let us note that
the infrared properties of the gluon   propagators have already been
used to study thermodynamics of Yang--Mills plasma at $T>T_{c}$
in the Coulomb gauge~\cite{ref:quasiparticles}. The idea is that
constraining  the configuration space through the gauge fixing condition
leads to a change in  the dispersion relations for gluons, or position of
the poles of the propagator. In the Landau gauge, a relation between the
non-perturbative Green's functions and the thermodynamic potential was
discussed  in Refs.~\cite{ref:hot:Landau,ref:hot:Landau:recent}.
Below we  concentrate on zeros rather than  poles of the propagators
and demonstrate that zeros give rise to new $T^4$ terms in the equation of state.

To illustrate the basic idea  consider first a toy model
describing one degree of freedom, or a real
scalar field $\phi(x)$ with a
quadratic  action:
\beqn
S[\phi] = - \frac{1}{2} \int_0^{1/T} \!\dd \tau \! \int \!\dd^3 x \, \phi(x) \,
\cD^{(-1)}(x-y) \, \phi(y)\,,
\label{eq:S}
\eeqn
where  the propagator $\cD(x-y)$ is defined as
\beqn
\cD(x-y) = \avr{\phi(x) \phi(y)}\,,
\label{eq:prop}
\eeqn
and we use the imaginary time formalism.
The partition function can be calculated in the standard way~\cite{ref:Kapusta}:
\beqn
\cZ = \int \dD \phi \, e^{-S^{(2)}[\phi]} = \exp\Bigl\{\frac{1}{2} \Tr \log (T^2 \cD)\Bigr\}\,,
\eeqn
where we omit the irrelevant prefactor.
Taking  the trace over all the states, one gets for the pressure~(\ref{eq:pressure:1}):
\beqn
P = - \frac{T}{2} \intp{p} \sum_{n \in \Z} \log \Bigl[T^2 \cD(\bp,p_4) \Bigr]{{\Biggl|}}_{p_4 = \omega_n}\,,
\label{eq:pressure:2}
\eeqn
where the sum runs over the Matsubara frequencies, $\omega_n = 2 \pi n T$.
The energy density is given by
\beqn
\varepsilon = - T \intp{p} \sum_{n \in \Z}
\frac{\partial \log [p^2_4 \, \cD(\bp,p_4)]}{\partial \log p^2_4}{{\Biggl|}}_{p_4 = \omega_n}\,.
\label{eq:energy:2}
\eeqn

The temperature dependent part of this sum can be evaluated following Refs.~\cite{ref:Kapusta,ref:Kislinger}:
\beqn
\varepsilon(T) = - \frac{1}{\pi i} \intp{p}
\hskip -2mm \int\limits^{+ i \infty + \epsilon}_{- i \infty + \epsilon} \hskip -2mm
\dd p_0 \, f_T(p_0)
\,
\frac{F(p_0) + F(-p_0)}{2}\ \
\label{eq:int:1}
\\
F_\varepsilon(p_0) = \frac{\partial \log [p^2_4 \, \cD(\bp,p_4)]}{\partial \log p^2_4}{{\Biggl|}}_{p_4 = - i p_0}
\hskip -10mm \,, \qquad\qquad
\label{eq:f:e}
\eeqn
where $f_T(p_0) = 1/(e^{p_0/T}{-}1)$
is the Bose-Einstein distribution for a particle with energy $\omega = p_0$.
Equation~\eq{eq:int:1} is valid provided the analytical function $F(p_0)$ does not have
poles at purely imaginary~$p_0$.
Since the integrand in Eq.~\eq{eq:int:1} is fast converging as $p_0 \to + \infty$, the
contour of integration can be closed as a semicircle in the right half of the $p_0$-complex
plane, reducing the integral~\eq{eq:int:1} to contribution of
poles of the function $F_\varepsilon(p_0)$.

Poles of the propagator $\cD$ become
poles of the function $F_\varepsilon(p_0)$   and contribute to $\epsilon(T)$.
This agrees with our intuition since the energy spectrum of the model~\eq{eq:S}
is determined by the poles of the propagator $\cD$.
However, the crucial point is that not only the poles of the propagator, but also its {\it zeros}
contribute to the energy density~\eq{eq:int:1}, \eq{eq:f:e}. In fact, zeros of the propagator
$\cD$ become poles of the function $F_\varepsilon(p_0)$ as well.

Imagine that the propagator~\eq{eq:prop} in the infrared region has the same  criticality
as the gluon or ghost propagators~\eq{eq:propagator:critical} in Yang--Mills theory:
\beqn
D^{(\gamma)} = {\mathrm{const}}\,{(p^2)}^\gamma/(p^2 + m^2)\,.
\label{eq:prop:anomalous}
\eeqn
Then we get from Eq.~\eq{eq:f:e} the expression
\beqn
F^{(\gamma)}_\varepsilon(p_0) = \frac{\omega_\bp^2}{\omega^2_\bp - p_0^2}
- \gamma \frac{p^2_0}{\bp^2 - p_0^2}\,,
\label{eq:f:e:gamma}
\eeqn
with two poles at the ${\mathrm{Re}} \, p_0 > 0$ half of the complex $p_0$--plane:
at $p_0 = \omega_\bp$ and at $p_0 = |\bp|$. The residues, respectively, are
$- \omega_\bp/2$ and $\gamma |\bp|/2$. The energy density~\eq{eq:energy:2}~is
\beqn
& & \varepsilon(T)
\label{eq:sl2}
= \varepsilon^{\mathrm{free}}(T,m) - \gamma \, \varepsilon^{\mathrm{free}}(T,0)\,, \\
& & \varepsilon^{\mathrm{free}}(T,m) = \intp{p} \, \omega_\bp \, f_T (\omega_\bp)\,.
\label{eq:sl}
\eeqn
Here $\varepsilon^{\mathrm{free}}$ is the energy density of free relativistic particles of mass $m$.

It is clear from Eq.~\eq{eq:sl2} that the infrared suppression (with $\gamma>0$) in the particle
propagator acts oppositely to the pole: the suppression ``subtracts'' {\em massless} degrees of freedom
(with the dispersion relation corresponding to a relativistic massless ``particle'' $\omega_\bp = |\bp|$\
and with $\varepsilon_{\mathrm{free}}(T,0)  \propto T^4$)
while a pole  ``supplies'' them to the spectrum. The same effect is also
from the expression for the pressure~\eq{eq:pressure:2} because
$\log D^{(\gamma)} = \gamma \log(p^2) - \log(p^2 + m^2)$,
up to a momentum--independent term.
Note that a zero of the propagator corresponds to a {\it negative}  pressure and  energy density.
Thus,   vanishing of  the propagator in the infrared with  critical exponent $\gamma$ acts as
a ``reservoir'' for $\gamma$ massless degrees of freedom where the quantity $\gamma$ is, in
general, non-integer.
As we see from our illustrative example, the low-temperature asymptotics of the thermodynamic
functions are determined by the infrared behaviour of the particle propagator.

Let us now turn back to Yang-Mills theory.
In the leading, one-loop approximation (in terms of  dressed propagators)
pressure in Yang--Mills theory at vanishingly low temperatures is given by
the expression~\eq{eq:pressure:2} with
\beqn
\cD(p^2) \to {\Biggl(\frac{D_L(p^2)D_T^2(p^2)[D_L(p^2) + D_T(p^2)]}{D_G^2(p^2)}\Biggr)}^{N^2_c-1} \hskip -5mm. \
\label{eq:D:YM:2}
\eeqn
The ghost structure function $D_G$ in the denominator of Eq.~\eq{eq:D:YM:2} corresponds to the Faddeev--Popov determinant
which subtracts unphysical degrees of freedom. Indeed, if one neglects the interactions,
$g=0$, then
\beqn
D^{\mathrm{free}}_L = D^{\mathrm{free}}_T = D_G^{\mathrm{free}} = 1/p^2\,,
\eeqn
and therefore the energy and pressure are given by Eq.~\eq{eq:SB} with $N_{d.f.} = 2 (N^2_c - 1)$, as expected.

However, if one takes into account the infrared dressing of the
propagators~\eq{eq:propagator:critical}, then
one gets for the pressure
\beqn
P & = & (2 - \gamma_L - 2 \gamma_T + 2 \gamma_G - \min[\gamma_L,\gamma_T])
\nonumber \\
& & \cdot (N^2_c - 1) C_{\SB} T^4 + \dots\,, \label{eq:pressure:3}
\eeqn where ellipsis denote subleading $O(T^\kappa)$ terms with
$\kappa > 4$, associated with cuts in the complex $p_0$-plane.
The suppression~\eq{eq:suppression} of thermodynamic quantities at
low temperatures implies a relation between the infrared exponents:
\beqn
\gamma_L + 2 \gamma_T - 2 \gamma_G + \min[\gamma_L,\gamma_T] - 2 = 0\,.
\label{eq:rel:T}
\eeqn
Since the functions $H_i$,
$i=L,T,G$, are regular at $p^2 = 0$ the poles and/or zeros of these
functions do not contribute to a $T^4$ term at
very
low
temperatures. Indeed, poles and/or zeros at  momenta
${\mathrm{Re}} \, p_0 > 0$ would correspond to massive--like
contributions which are exponentially
suppressed by the Bose-Einstein factor as $T \to 0$.

Coming back to the question of gauge dependence of our results, we mention that the form
of the propagators~\eq{eq:propagator:critical} assumes local Lorentz isotropy of the gauge condition
at zero temperature. However, in a general case (like, for example, in the Coulomb gauge) or/and
at a finite temperature (as we discuss in detail below) the form of the propagator is not rotationally
invariant, and the r\^ole of the infrared critical exponent is played by the degree $\gamma$ of the zero,
corresponding to the free dispersion relation $p^{\mathrm{free}}_0 = |\bp|$. In the Coulomb gauge
effects of the Fundamental Modular region lead to a substantial modification of the gluon dispersion
relation~\cite{ref:quasiparticles}. Namely, relation between the energy and momentum of gluon is
no longer linear at low momenta.  Thus, the gluon's contribution to the equation of state in this gauge
does not contain $T^4$ terms in the limit of low temperature, in agreement with lattice
simulations~\cite{ref:quasiparticles}.

The relation~\eq{eq:rel:T} provides a new constraint on the
critical exponents. Existing lattice data does not allow to check it directly.
One can try to unify (\ref{eq:pressure:3}) with other relations suggested in the
literature. Assuming  validity of (\ref {eq:KugoOjima})
and $\gamma_L=\gamma_T$   we get a system of constraints which allows to uniquely
fix all the exponents and get what we would call simplified solution, $\gamma_D = 2/5$ and $\gamma_G = -1/5$.

The positivity of $\gamma_D$, exhibited by this solution,
is consistent with the infrared suppression of
the gluon propagator expected on other grounds ~\cite{ref:infrared:review}.
Also, the ghost exponent is  consistent with the numerical result of Ref.~\cite{ref:Michael}.
However, for the simplified solution the gluon infrared exponent has too small value to
guarantee vanishing of the gluon propagator at $p=0$. Also, the simplified solution is in
variance with results of Refs.~\cite{ref:Dyson:Schwinger,ref:large:volume2}.
One of the reasons could be violation at the presently available lattices
of the equality $\gamma_L=\gamma_T$ which is granted at zero temperature.
Indeed, according to numerical results of Ref. \cite{ref:hot:Landau:recent}
the transverse-longitudinal degeneracy is not reached yet
even at the lowest non-zero temperatures available now.
One might guess that the discrepancy between the simplified solution
and the existing simulations is rooted in the fact that the lowest
currently available non-zero temperatures are still too high to be
compared with our analytical predictions.

One can ask a question whether   the relation~\eq{eq:rel:T}
between the critical exponents is affected by the presence of quarks.
As far as the quarks are massive both
numerical~\cite{ref:quark:numer} and analytical
results~\cite{ref:quark:analyt} indicate absence of the infrared
particularities (zeros or poles) in the quark structure functions in
the Landau gauge. Consequently our relation~\eq{eq:rel:T} is
insensitive to the presence of the massive quarks.
This conclusion agrees nicely with the result of Ref.~\cite{ref:quark:analyt}.
However, for {\it exactly} massless quarks the $T^4$-terms  do
appear in the thermodynamical quantities in the $T\to 0$ limit
because of the presence of {\it exactly} massless Nambu-Goldstone
particles in the spectrum. Then the right hand side
Eq.~\eq{eq:rel:T} should be equal to $-N_{\mathrm{NG}}/(N_c^2 - 1)$,
where $N_{\mathrm{NG}}$ is the number of the Nambu-Goldstone bosons.

To summarize, the zeros in the momentum-space Green's function  are  as important
for the  thermodynamics as the poles. Although  zeros do not correspond to any states
in the spectrum  they contribute to the equation of state. The contribution of zeros
in the gluon propagator to coefficients in front of $T^{4}$ terms in the energy density
and pressure is negative. This observation can be interpreted in such a way that zeros
correspond to confinement, or binding of the originally massless gluons into massive glueballs.
Zeros of the ghost propagator, on the contrary, effectively supply massless degrees of
freedom into the equation of state. Moreover, the very existence of the mass gap in confining
theories requires vanishing of the overall contribution of massless degrees of
freedom from the low-temperature equation of state.
Hence, there arises the constraint \eq{eq:rel:T}  on the infrared exponents.
Confinement resolves itself into a new kind of analyticity.

\begin{acknowledgments}
We are grateful to A.~Di~Giacomo for reading the manuscript and comments,
to E.-M.~Ilgenfritz, A.~Maas and T.~Suzuki for communications,
and to G.~E.~Volovik for bringing to our attention Refs.~\cite{ref:Volovik,ref:zeroes}.
We are thankful to the members of Institute for Theoretical Physics of Kanazawa
University and Research Institute for Information Science and Education of Hiroshima University
for hospitality and stimulating environment.
This work was partly supported by Federal Program of the Russian Ministry of Industry, Science
and Technology No. 40.052.1.1.1112.
M.N.Ch is supported by the JSPS grant No. L-06514, by
Grant-in-Aid for Scientific Research by Monbu-kagakusyo, No. 13135216 and by a CNRS grant.
\end{acknowledgments}

\end{document}